\documentclass[showpacs,twocolumn,preprintnumbers,amsmath,amssymb]{revtex4}
\usepackage{bbm}
\usepackage{amsfonts}

\usepackage{graphicx}
\usepackage{amssymb}   
\usepackage{dcolumn}
\usepackage{bm}
\usepackage[colorlinks,citecolor=blue, linkcolor=blue,hyperindex]{hyperref}
\usepackage{url}
\begin{document}

\vspace{2.5cm}
\title{Protecting qutrit-qutrit entanglement by weak measurement and reversal}
\author{Xing Xiao$^{1,2}$}
\altaffiliation{Corresponding author: xiaoxing1121@gmail.com}
\author{Yan-Ling Li$^{3}$}
\affiliation{${1}$ College of Physics and Electronic Information,
Gannan Normal University, Ganzhou 341000, China \\
${2}$ Institute of Optoelectronic Materials and Technology, Gannan
Normal University, Ganzhou 341000, China \\
${3}$ School of Information Engineering, Jiangxi University of
Science and Technology, Ganzhou 341000, China}

\begin{abstract}
Entangled states in high dimensional systems are of great interest
due to the extended possibilities they provide in quantum
information processing. Recently, Sun [Phys. Rev. A 82, 052323
(2010)]  and Kim [Nat. Phys. 8, 117 (2012)] pointed out that weak
measurement and quantum weak measurement reversal can actively
combat decoherence. We generalize their studies from
qubits to qutrits under amplitude damping decoherence. We find that
the qutrit-qutrit entanglement can be partially retrieved for
certain initial states when only weak measurement reversals are
performed. However, we can completely defeat amplitude damping
decoherence for any initial states by the combination of prior weak
measurements and post optimal weak measurement reversals. The
experimental feasibility of our schemes is also discussed.
\end{abstract}
\maketitle

\section{Introduction}
Quantum entanglement is not only a remarkable characteristic which
distinguishes the quantum realm from the classical one, but also a key
resource for quantum information and quantum computation
\cite{nielsen}. However, in realistic quantum information
processing, entanglement is inevitably affected by the interaction
between the system and its environment, which leads to degradation and,
in certain cases, entanglement sudden death (ESD)
\cite{yuting1,yuting2,almeida}. Thus, it is very important to
protect entanglement from environmental noise.

Weak measurements \cite{note} are generalizations of von
Neumann measurements and are associated with a positive-operator
valued measure (POVM). For weak measurements \cite{koro,koro2}, the
information extracted from the quantum system is deliberately limited,
thereby keeping the measured system's state from randomly collapsing towards
an eigenstate. Thus, it would be possible to reverse the
initial state with some operations. Recently, it was pointed out
that weak measurements and quantum weak measurement reversals can
effectively protect the quantum states of a single qubit system from
decoherence \cite{sunqq1,koro1,xiao}; this idea has also been
extended to protect the entanglement of two-qubit systems
\cite{sunqq2,yskim,manzx,liyl} from amplitude damping decoherence.
Until now, probabilistic reversal with a weak measurement has
already been experimentally demonstrated on a superconducting phase
qubit \cite{katz}, as well as on a photonic qubit
\cite{yskim,yskim2}.

Most studies of weak measurements concerning the protection of
entanglement are restricted to two dimensional (2D) systems.
However, quantum information tasks require high dimensional
bipartite entanglement. It is well known that high dimensional
entangled systems such as qutrits \cite{mair,GMT,inoue} can offer
significant advantages for the manipulation of information carriers.
For instance, biphotonic qutrit-qutrit entanglement \cite{BPL}
enables more efficient use of communication channels \cite{SPW}.
Moreover, high dimensional entangled systems offer higher
information-density coding and greater resilience to errors than 2D
entangled systems in quantum cryptography \cite{GMN}. However,
practical applications of such protocols are only conceivable when
the prepared high dimensional entangled states have sufficiently
long coherence times for manipulation.

In this paper, we propose using weak measurements to preserve the
entanglement of two initially entangled qutrits which suffer
independent amplitude damping noise. Our schemes for protecting
entanglement are based on the fact that weak quantum measurement can
be reversed probabilisticlly. We specifically consider two simple
schemes as shown in Fig.~1. Similar schemes have been discussed only in one  
or two-qubit systems \cite{yskim, yskim2,sunqq2}, while we consider a
qutrit-qutrit version in this paper. The first scheme is
``\emph{amplitude damping} + \emph{weak measurement reversal}''. In
this case, unlike the entanglement decaying exponentially to zero in
amplitude damping decoherence, we show that the weak measurement
reversal procedure partially recovers the entanglement under most
conditions. The limitation of this scheme is that ESD still occurs
in some particular situations. As an improvement on the former, the
second scheme is ``\emph{weak measurement} + \emph{amplitude
damping} + \emph{weak measurement reversal}''. In this case, we find
the combination of prior weak measurement and post weak measurement
reversal can actively combat decoherence. Moreover, it can
effectively circumvent ESD. The physical mechanism of the second
scheme is that a prior weak measurement intentionally moves each
qutrit close to its ground state. The amplitude damping decoherence
is naturally suppressed in this `lethargic' state, and the
entanglement is therefore preserved \cite{ank}.

This paper is organized as follows: In Section~\ref{s2}, we
introduce amplitude damping noise operators for the qutrit case, then we
generalize the weak measurement and weak measurement reversal
operators from qubit to qutrit. In Section~\ref{s3}, we propose two
different schemes to protect qutrit-qutrit entanglement. In Section~\ref{s4},
we give a brief discussion of the experimental feasibility
of our schemes. Finally, we summarize our conclusions in Section~\ref{s5}.

\section{\label{s2}Basic theory}

\subsection{amplitude damping for qutrits}
The amplitude damping noise is a prototype model of a dissipative
interaction between a quantum system and its environment
\cite{nielsen}. For example, the amplitude damping noise model can
be applied to describe the spontaneous emission of a photon by a
two-level system into an environment of photon or phonon modes at
zero (or very low) temperature in (usually) the weak Born-Markov
approximation.
 \begin{figure}[ht]
 \label{f1}
   \centering
   \includegraphics[width=3in,height=2.4in]{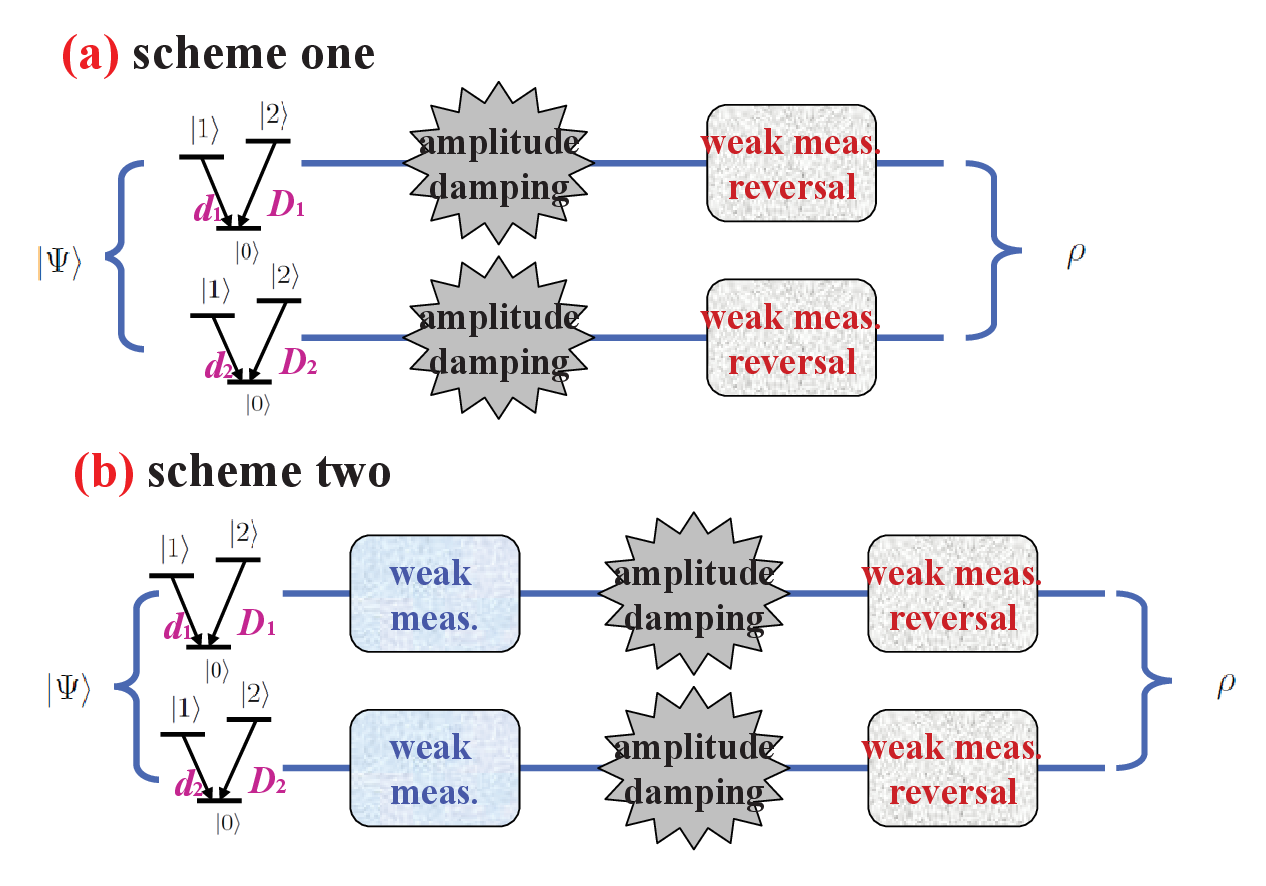}
   \caption{\label{FirstFigure} (color online) Schemes for protecting entanglement from decoherence using
     weak measurement and weak measurement reversal: (a), Two entangled
     qutrits go through independent amplitude damping channels and then a
     weak measurement reversal is performed on each qutrit. (b),
     Similar to (a), but a weak measurement is applied before each
     qutrit undergoes decoherence.}
 \end{figure}

For qutrits, the situations are more complicated as there are three
configurations of the 3-level system to be taken into account
\cite{hioe}. Here, we will focus on the so called V-configuration.
We denote the lower level as $|0\rangle$ and the two upper levels as
$|1\rangle$ and $|2\rangle$, respectively. We assume that only
dipole transitions between levels $|1\rangle\rightarrow|0\rangle$
and $|2\rangle\rightarrow|0\rangle$ are allowed. If the environment
is in a vacuum state, the amplitude damping noise which corresponds
to the spontaneous emission from the V-configuration qutrit can be
represented by the following map \cite{check}:
\begin{eqnarray}
\label{e1}
&&|0\rangle_{S}|0\rangle_{E}\rightarrow|0\rangle_{S}|0\rangle_{E},\nonumber\\
&&|1\rangle_{S}|0\rangle_{E}\rightarrow\sqrt{1-d}|1\rangle_{S}|0\rangle_{E}+\sqrt{d}|0\rangle_{S}|1\rangle_{E},\\
&&|2\rangle_{S}|0\rangle_{E}\rightarrow\sqrt{1-D}|2\rangle_{S}|0\rangle_{E}+\sqrt{D}|0\rangle_{S}|1\rangle_{E},\nonumber
\end{eqnarray}
where $d,D\in[0,1]$ represents the decay rates of the upper levels
$|1\rangle$ and $|2\rangle$, respectively.

\subsection{weak measurement for qutrits}
The null-result weak measurement that we consider is the POVM or
partial-collapse measurement originally discussed in Refs.
\cite{koro,koro2}. It is different from amplitude damping in the
sense that we add an ideal detector to monitor the environment
function as follows: the detector clicks with a probability \emph{p}
if there is an excitation in the environment and never clicks with a
probability $1-p$ if no excitation is detected in the environment.
For the qutrit case, we can construct the POVM elements as:
$M_{1}=diag(0,\sqrt{p},0)$, $M_{2}=diag(0,0,\sqrt{q})$ and
$M_{3}=diag(1,\sqrt{1-p},\sqrt{1-q})$, where $p$ and $q$ represent
the weak measurement strengths of transitions
$|1\rangle\rightarrow|0\rangle$ and $|2\rangle\rightarrow|0\rangle$,
respectively. The measurement operators $M_{1}$ and $M_{2}$ are
identical to the normal projection measurements in which the state
of the qutrit is irrevocably collapsed to the ground state and an
excitation is emitted from system to environment. They are not
reversible and we therefore discard the result from experiments which produced clicks, thereby
removing the terms $\sqrt{p}|0\rangle_{S}|1\rangle_{E}$ and
$\sqrt{q}|0\rangle_{S}|1\rangle_{E}$ from the state map.
Fortunately, the measurement operator $M_{3}$ is a weak (or
partial-collapse) measurement for the single qutrit that we are
interested in in this paper. We rewrite $M_{3}$ as
\begin{eqnarray}
\label{e2}
&&|0\rangle_{S}|0\rangle_{E}\rightarrow|0\rangle_{S}|0\rangle_{E},\nonumber\\
&&|1\rangle_{S}|0\rangle_{E}\rightarrow\sqrt{1-p}|1\rangle_{S}|0\rangle_{E},\\
&&|2\rangle_{S}|0\rangle_{E}\rightarrow\sqrt{1-q}|2\rangle_{S}|0\rangle_{E},\nonumber
\end{eqnarray}

\subsection{weak measurement reversal for qutrits}
Except for von Neumann projective measurements, any weak or
partial-collapse measurement could be reversed \cite{YWC}. According
to Ref.~\cite{YWC}, it is easy to construct the reversed weak
measurement operator of the null-result weak measurement as shown in
Eq.~(\ref{e2}). The single-qutrit reversing measurement ($M_r$) is
also a non-unitary operation that can be written as
\begin{eqnarray}
\label{e3} M_{r}=\left(
  \begin{array}{ccc}
    \sqrt{(1-p_r)(1-q_r)} & 0 & 0 \\
    0 & \sqrt{1-q_r} & 0 \\
    0 & 0 & \sqrt{1-p_r} \\
  \end{array}
\right),
\end{eqnarray}
where $p_r$ and $q_r$ are the strengths of the reversing
measurements. As the matrix is non-unitary, the probability of
successful reversal will always be less than unity.

\section{\label{s3}Protection of qutrit-qutrit entanglement}

\subsection{scheme one}
We first check the efficiency of the first scheme as shown in Fig.~1(a).
For simplicity, we assume two identical qutrits are initially
prepared in the following state
\begin{equation}
\label{e4}
|\Psi\rangle=\alpha|00\rangle+\beta|11\rangle+\gamma|22\rangle,
\end{equation}
where $|\alpha|^{2}+|\beta|^{2}+|\gamma|^{2}=1$. Such a
qutrit-qutrit entangled state can be experimentally prepared by
utilizing the orbital angular momentum of photons \cite{mair,inoue}
. We assume they suffer independent but identical amplitude damping
noise (i.e., $d_{1}=d_{2}=D_{1}=D_{2}=D$). Then the initial pure
state inevitably evolves into a mixed state in the presence of
noise.
\begin{equation}
\label{e5}
  \rho_{d}=\sum_{i=1}^{9}\varepsilon_{i}|\Psi\rangle\langle\Psi|\varepsilon_{i}^{\dagger},
\end{equation}
where $\varepsilon_{i}=E_{j}\otimes E_{k}, (j,k=0,1,2)$ are the
Kraus operators. In the standard product basis
$\{|j,k\rangle=|3j+k+1\rangle\}$, the non-zero elements of
$\rho_{d}$ are:
\begin{eqnarray}
\label{e6}
\rho_{11}&=&|\alpha|^{2}+D^{2}(|\beta|^{2}+|\gamma|^{2}),\nonumber\\
\rho_{15}&=&\rho_{51}^{*}=(1-D)\alpha\beta^{*},\nonumber\\
\rho_{19}&=&\rho_{91}^{*}=(1-D)\alpha\gamma^{*},\\
\rho_{22}&=&\rho_{44}=D(1-D)|\beta|^{2},\nonumber\\
\rho_{33}&=&\rho_{77}=D(1-D)|\gamma|^{2},\nonumber\\
\rho_{55}&=&(1-D)^{2}|\beta|^{2},\nonumber\\
\rho_{59}&=&\rho_{95}^{*}=(1-D)^{2}\beta\gamma^{*},\nonumber\\
\rho_{99}&=&(1-D)^{2}|\gamma|^{2},\nonumber
\end{eqnarray}

After the amplitude damping decoherence, we perform quantum
measurement reversal operations on each qutrit as shown in Eq.~(\ref{e3}).
The non-zero elements of the final reduced density matrix
$\rho_{r}$ are:
\begin{eqnarray}
\label{e7}
\rho_{11}&=&[(1-p_{r})^{2}|\alpha|^{2}+D^{2}(1-p_{r})^{2}(|\beta|^{2}+|\gamma|^{2})]/C_{1},\nonumber\\
\rho_{15}&=&\rho_{51}^{*}=(1-D)(1-p_{r})\alpha\beta^{*}/C_{1},\nonumber\\
\rho_{19}&=&\rho_{91}^{*}=(1-D)(1-p_{r})\alpha\gamma^{*}/C_{1},\\
\rho_{22}&=&\rho_{44}=D(1-D)(1-p_{r})|\beta|^{2}/C_{1},\nonumber\\
\rho_{33}&=&\rho_{77}=D(1-D)(1-p_{r})|\gamma|^{2}/C_{1},\nonumber\\
\rho_{55}&=&(1-D)^{2}|\beta|^{2}/C_{1},\nonumber\\
\rho_{59}&=&\rho_{95}^{*}=(1-D)^{2}\beta\gamma^{*}/C_{1},\nonumber\\
\rho_{99}&=&(1-D)^{2}|\gamma|^{2}/C_{1},\nonumber
\end{eqnarray}
where
$C_{1}=(1-p_{r})^{2}|\alpha|^{2}+[(1-D)^{2}+2D(1-D)(1-p_{r})+D^2(1-p_{r})^{2}](|\beta|^{2}+|\gamma|^{2})$
is the normalization parameter.

To quantify the qutrit-qutrit entanglement change under amplitude
damping noise and weak measurement reversal, we need an effective
measure of mixed qutrit state entanglement since damping causes the
pure states to evolve into mixed states. One usually takes the
entanglement of formation \cite{bennet} as such a measure, but in
practice it is not known how to compute this measure for mixed
states of $d\otimes d$ dimensional systems in the case when $d>2$. A
computable measure of distillable entanglement of mixed states was
proposed in Ref.~\cite{vidal}. It is based on the trace norm of the
partial transposition $\rho^{T}$ of the state $\rho$. From the
Peres' criterion of separability \cite{peres}, it follows that if
$\rho^{T}$ is not positive, then $\rho$ is entangled. Hence one
defines the negativity of the state $\rho$ as
\begin{equation}
\label{e8}
  N=\frac{||\rho^{T}||-1}{2}.
\end{equation}
$N$ is equal to the absolute value of the sum of negative eigenvalues
of $\rho^{T}$ and is an entanglement monotone \cite{vidal}, but it
cannot detect bound entangled states \cite{horo}.
\centerline{\includegraphics[width=3in,height=5.2in]{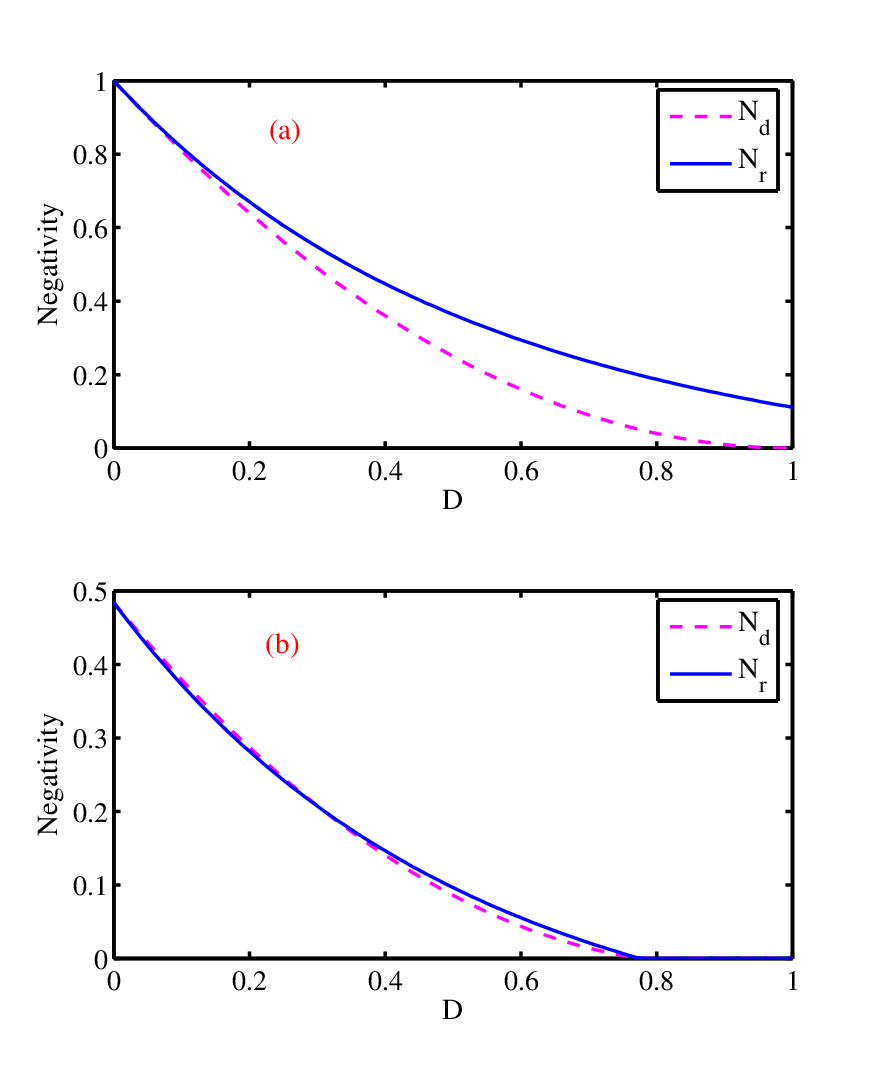}}
\begin{center}
\parbox{8.5cm}{\small{\bf Fig.2.} (color online) Negativity as a function of $D$, where we have
chosen the optimal reversing measurement strength to be $p_{r}=D$.
(a) $|\Psi\rangle=1/\sqrt{3}(|00\rangle+|11\rangle+|22\rangle)$, the
reversed negativity is always higher than $N_{d}$. When
$p_{r}\rightarrow1$, $N_{d}$ goes to 0 while $N_{r}$ is finite. (b)
$|\Psi\rangle=\sqrt{3/8}|00\rangle+\sqrt{5/8}|11\rangle+0|22\rangle$,
$N_{r}$ is not always higher than $N_{d}$, and both $N_{d}$ and
$N_{r}$ go to 0 when ESD appears.}
\end{center}

According to Eq.~(\ref{e8}), it is easy to calculate the damped
negativity ($N_d$) and the reversed negativity ($N_r$). However, the
general analytic expressions of negativity for $\rho_{d}$ and
$\rho_r$ are too complicated to present as they depend on the
relationships between the initial parameters $\alpha$, $\beta$,
$\gamma$ and the decoherence parameter $D$. Hence we will present
numerical results in the discussions below. In Fig.~2, we show how
$N_{d}$ and $N_{r}$ behave for two particular initial states under
amplitude damping decoherence and corresponding optimal reversing
measurements. We have chosen the optimal reversing measurement
strength to be $p_{r}=D$ which gives the maximum amount of
entanglement of the two-qutrit state $\rho_{r}$. For
$|\Psi\rangle=1/\sqrt{3}(|00\rangle+|11\rangle+|22\rangle)$, we note
that the damped negativity $N_{d}$ decays as the
decoherence strength \emph{D} increases, while the reversed negativity $N_{r}$
approaches a finite value. The reversed negativity $N_{r}$ is
higher than the negativity $N_{d}$ regardless of the
decoherence strength parameter, as shown in Fig.~2(a). However, for
$|\Psi\rangle=\sqrt{3/8}|00\rangle+\sqrt{5/8}|11\rangle+0|22\rangle$,
we find the reversed negativity $N_{r}$ is not always higher than
the negativity $N_{d}$. Moreover, the reversed negativity $N_{r}$
suffers sudden death as well as $N_{d}$, as shown in Fig.~2(b). The
reason is straightforward as all operations are local, and no
entanglement can be created between two independent qutrits in a
separable state. The above results for qutrits are quite in
accordance with those for qubits discussed in Ref.~\cite{sunqq2}.

As the weak measurement reversals are non-unitary operations, this
scheme naturally has less than unity success probability. Under the
optimal reversing weak measurements (i.e., $p_{r}=D$), the
corresponding success probability is:
\begin{equation}
\mathcal
{P}_{1}=(1-D)^{2}\left[1+(|\beta|^2+|\gamma|^2)(2D+D^2)\right].
\end{equation}
It is clear that $\mathcal{P}_{1}\rightarrow0$ when $D\rightarrow1$.

\subsection{scheme two}

As shown above, the first scheme has some limitations regarding the
protection of entanglement and circumvention of ESD. In this
section, we show that an improved scheme first proposed by Kim
\emph{et al.} \cite{yskim} can completely circumvent the decoherence and
protect the qutrit-qutrit entanglement even if ESD occurs. The key
difference is that a prior weak measurement is applied on each
qutrit before it suffers amplitude damping decoherence, as depicted
in Fig.~1(b).

The whole procedure is as follows: for each qutrit, first a prior
weak measurement with strength $p$ is performed, then it goes
through the amplitude damping channel, and finally a post weak
measurement reversal with strength $p_{r}$ is carried out. After
these operations, the non-zero elements of the reduced density
matrix $\rho_{wr}$ are:
\begin{eqnarray}
\rho_{11}&=&[(1-p_{r})^{2}|\alpha|^{2}+(1-p)^{2}D^{2}(1-p_{r})^{2}\nonumber\\
&&(|\beta|^{2}+|\gamma|^{2})]/C_{2},\nonumber\\
\rho_{15}&=&\rho_{51}^{*}=(1-p)(1-D)(1-p_{r})\alpha\beta^{*}/C_{2},\nonumber\\
\rho_{19}&=&\rho_{91}^{*}=(1-p)(1-D)(1-p_{r})\alpha\gamma^{*}/C_{2},\\
\rho_{22}&=&\rho_{44}=(1-p)^{2}D(1-D)(1-p_{r})|\beta|^{2}/C_{2},\nonumber\\
\rho_{33}&=&\rho_{77}=(1-p)^{2}D(1-D)(1-p_{r})|\gamma|^{2}/C_{2},\nonumber\\
\rho_{55}&=&(1-p)^{2}(1-D)^{2}|\beta|^{2}/C_{2},\nonumber\\
\rho_{59}&=&\rho_{95}^{*}=(1-p)^{2}(1-D)^{2}\beta\gamma^{*}/C_{2},\nonumber\\
\rho_{99}&=&(1-p)^{2}(1-D)^{2}|\gamma|^{2}/C_{2},\nonumber
\end{eqnarray}
where
$C_{2}=(1-p_{r})^{2}|\alpha|^{2}+(1-p)^{2}[(1-D)^{2}+2D(1-D)(1-p_{r})+D^2(1-p_{r})^{2}](|\beta|^{2}+|\gamma|^{2})$.

\centerline{\includegraphics[width=3in,height=5.2in]{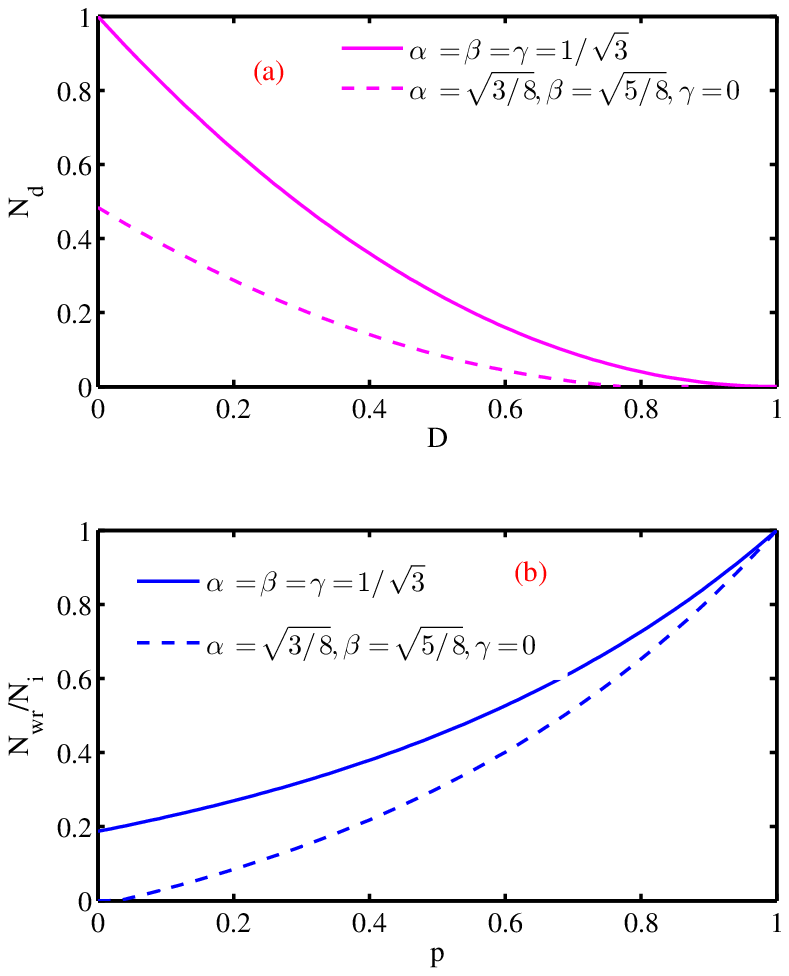}}
\begin{center}
\parbox{8.5cm}{\small{\bf Fig.3.} (color online) For two particular initially entangled qutrit-qutrit states
$|\Psi\rangle=1/\sqrt{3}(|00\rangle+|11\rangle+|22\rangle)$ (solid
line) and
$|\Psi\rangle=\sqrt{3/8}|00\rangle+\sqrt{5/8}|11\rangle+0|22\rangle$
(dashed line): (a) Negativity $N_{d}$ as a function of decoherence
strength $D$. (b) The ratio of $N_{wr}$ to $N_{i}$ as a function of
weak measurement strength \emph{p} when $D=0.8$ and an optimal
reversing measurement is performed.}
\end{center}

Following the methods demonstrated in Refs.~\cite{koro1,yskim}, the
optimal reversing measurement strength that gives the maximum amount
of entanglement for the two-qutrit state $\rho_{wr}$ is calculated to
be $p_{r}=p+D\bar{p}$ where $\bar{p}=1-p$. We still consider the
same initial states in scheme one and compare the effectiveness of
these two schemes for suppressing amplitude damping noise. In Fig.~3,
we show the protection of entanglement from decoherence by using
a weak measurement and weak measurement reversal. As we know, the
qutrit-qutrit entanglement decays monotonously with increasing
\emph{D} and even experiences ESD when
$|\Psi\rangle=\sqrt{3/8}|00\rangle+\sqrt{5/8}|11\rangle+0|22\rangle$.

However, it is clear that the qutrit-qutrit entanglement can be
protected by the combined action of prior weak measurements and post
weak measurement reversals. In Fig.~3(a), we note that the 
negativity of
$|\Psi\rangle=1/\sqrt{3}(|00\rangle+|11\rangle+|22\rangle)$ is 0.04
when $D=0.8$, but it can be completely reversed to its initial
entanglement as $p\rightarrow1$ in Fig.~3(b). In Fig.~3(b), we have
introduced the ratio of $N_{wr}$ (negativity after the sequence of
weak measurement, decoherence and reversing measurement) to the
initial negativity $N_{i}$ to highlight the entanglement recovery efficiency.
To demonstrate the scheme's ability to circumvent ESD,
we choose $D=0.8$, at which point ESD appears for the initial state
$|\Psi\rangle=\sqrt{3/8}|00\rangle+\sqrt{5/8}|11\rangle+0|22\rangle$
in Fig.~3(a). We find the entanglement can be completely recovered
with a certain probability by the sequence of weak measurement and
weak measurement reversal, which is similar to that in Ref.~\cite{yskim}
where a two-qubit entangled state is considered.

Similarly to the first scheme, the success probability under the
optimal reversing weak measurements (i.e., $p_{r}=p+D\bar{p}$) can
be written as
\begin{equation}
\mathcal
{P}_{2}=(1-D)^{2}\bar{p}^{2}\left[1+(|\beta|^2+|\gamma|^2)(2D\bar{p}+D^{2}\bar{p}^{2})\right].
\end{equation}
We observe that for $p\rightarrow1$, the success probability
$\mathcal{P}_{2}\rightarrow0$ because the prior weak measurement is
reduced to an unrecoverable von Neumann projective measurement.

By comparing the two schemes, it is easy to find that the second
scheme is much more efficient than the first scheme at protecting
entanglement and circumventing ESD. Physically, this can be
explained as follows: From Equation~(\ref{e2}), we
know that the stronger the weak measurement strength $p$, the closer the
initial qutrit is reversed towards the $|0\rangle$ state. Once the
system is in $|0\rangle$, then it will be immune to amplitude
damping decoherence. In the first scheme, no prior weak measurement
is carried out before the qutrits go through the amplitude damping
channel, thus the amount of reversed entanglement highly depends on
the initial states and the decoherence strength $D$. In the second
scheme, prior weak measurements are performed to move the state
towards the $|00\rangle$ state, which does not experience amplitude damping
decoherence. Then optimal weak measurement reversals are performed
to revert the qutrits back to the initial state. Therefore, the
amount of reversed entanglement is not related to the decoherence
strength $D$ but depends on the weak measurement strength $p$.
Initial entanglement can entirely be recovered for any initial
state by the combined prior weak measurements and post weak
measurement reversals when $p\rightarrow1$.

\subsection{Discussions}
In the above analyses, we have assumed that the two qutrits are
identical and the decoherence parameters $d$ and $D$ are the same
for states $|1\rangle$ and $|2\rangle$. In fact, these two
schemes are universal for the most general case (i.e., $d_{1}\neq
d_{2}\neq D_{1}\neq D_{2}$). Following the same calculation
procedure as above, we plot in Fig.~4 the numerical results for the
two weak measurement schemes against the amplitude damping
decoherence. For the first scheme, the optimal reversing measurement
strength is calculated to be $p_{r_k}=d_{k}$ and $q_{r_k}=D_{k}$
($k=1,2$). Similarly, the optimal reversing measurement strength
should be $p_{r_k}=p_{k}+d_{k}\bar{p}_{k}$ and
$q_{r_k}=q_{k}+D_{k}\bar{q}_{k}$ for the second scheme.

\centerline{\includegraphics[width=3in,height=5.2in]{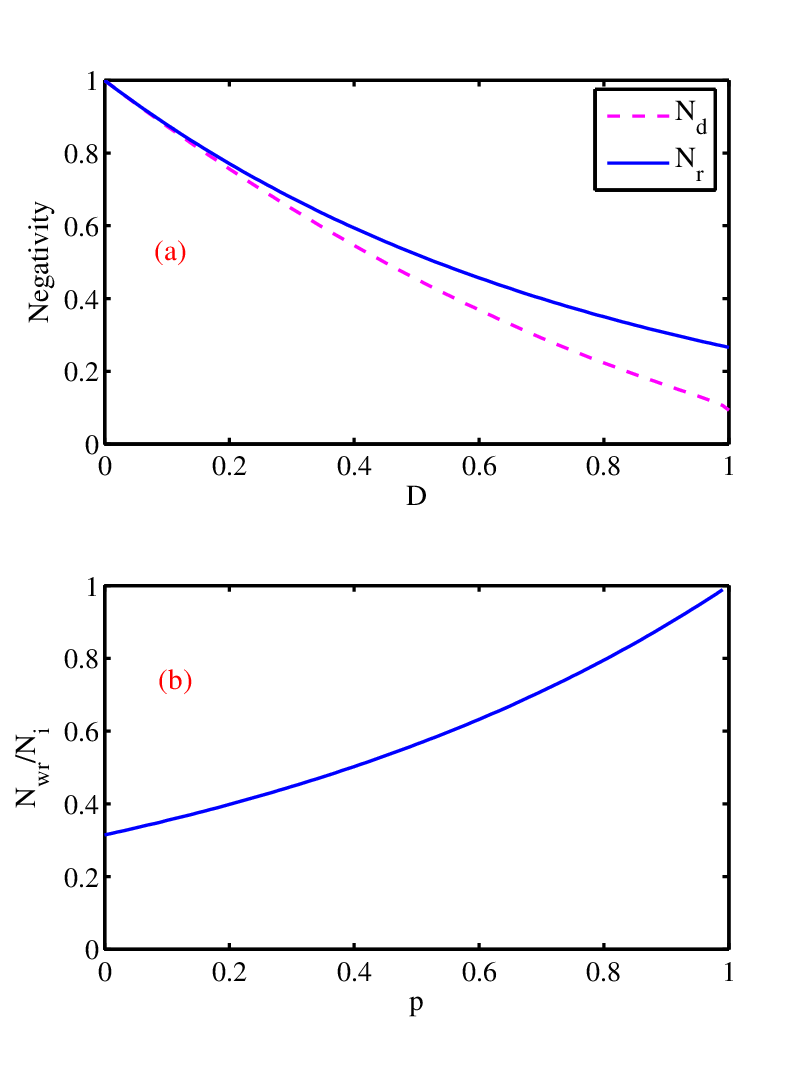}}
\begin{center}
\parbox{8.5cm}{\small{\bf Fig.4.} (color online) Entanglement
protection of state
$|\Psi\rangle=1/\sqrt{3}(|00\rangle+|11\rangle+|22\rangle)$ by weak
measurement and weak measurement reversal: (a) scheme one with the
decoherence parameters $d_{1}=D$, $d_{2}=0.7D$, $D_{1}=0.3D$ and
$D_{2}=0.6D$, (b) scheme two with the decoherence parameters
$d_{1}=0.8$, $d_{2}=0.5$, $D_{1}=0.4$ and $D_{2}=0.6$.}
\end{center}

\section{\label{s4}Experimental feasibility}
It is necessary to give a brief discussion of some key problems
which are related to the experimental implementation of our
procedure. Here, we restrict our discussions to the cavity QED
system which we think is the best candidate for the
experimental realization of our scheme.

\emph{Initial state preparation.} The V-type qutrit-qutrit
entanglement of Eq.~(\ref{e4}) could be generated by sending a pair
of momentum and polarization-entangled photons to two spatially
separated cavities in which a V-type atom is trapped \cite{lloyd}.
For the atomic level structure, we can take $^{87}Rb$ as our choice.
The state $|0\rangle$ corresponds to $|F=1, m_{F}=0\rangle$ of
$5^{2}S_{1/2}$, while the states $|1\rangle$ and $|2\rangle$
correspond to the degenerate levels $|F=1, m_{F}=1\rangle$ and
$|F=1, m_{F}=-1\rangle$ of $5^{2}P_{1/2}$, respectively. The
transitions $|1\rangle\rightarrow|0\rangle$ and
$|2\rangle\rightarrow|0\rangle$ emit right-circularly and
left-circularly polarized photons, so we can distinguish the
parameters $p$ and $q$ during the weak measurement.

\emph{Amplitude damping decoherence.} In a cavity QED system, the
amplitude damping decoherence is the natural spontaneous emission of
a photon from the excited state of an atom to its ground state. The
dynamic map of Eq.~(\ref{e1}) describes a dissipative interaction
between a V-type qutrit and its vacuum environment \cite{check}.

\emph{Weak measurement.} As shown in Sect.~\ref{s2}, we note that the only
difference between the AD decoherence map Eq.~(\ref{e1}) and weak
measurement map Eq.~(\ref{e2}) is the inclusion of the
$\sqrt{p}|0\rangle_S|1\rangle_E$ and
$\sqrt{q}|0\rangle_S|1\rangle_E$ terms. In this sense, we can add an ideal
single-photon detector to monitor the cavity. Whenever there is a
detector click, we discard the result. This postselection removes
the $\sqrt{p}|0\rangle_S|1\rangle_E$ and
$\sqrt{q}|0\rangle_S|1\rangle_E$ terms and hence a null-result weak
measurement is implemented.

\emph{Weak measurement reversal.} To reverse the effect of the weak
measurement ($M_{3}$), we only need to apply the inverse of $M_{3}$
\begin{equation}
M_{3}^{-1}=\left(
             \begin{array}{ccc}
               1 & 0 & 0 \\
               0 & \frac{1}{\sqrt{1-p}} & 0 \\
               0 & 0 & \frac{1}{\sqrt{1-q}} \\
             \end{array}
           \right)
\end{equation}
since $M_{3}^{-1}$ can be re-written as
\begin{eqnarray}
M_{3}^{-1}&&=\frac{1}{\sqrt{(1-p)(1-q)}}\mathcal{F}M_{3}\mathcal{F}M_{3}\mathcal{F}\nonumber\\
&&=\frac{1}{\sqrt{(1-p)(1-q)}}M_{r},
\end{eqnarray}
where $\mathcal{F}$ is the trit-flip operation
\begin{equation}
\mathcal{F}=\left(
             \begin{array}{ccc}
               0 & 0 & 1 \\
               1 & 0 & 0 \\
               0 & 1 & 0 \\
             \end{array}
           \right).
\end{equation}
Thus, the weak measurement reversal procedure of Eq.~(\ref{e3}) can
be constructed by the following five sequential operations on each
system qutrit: trit-flip ($\mathcal{F}$), weak measurement
($M_{3}$), trit-flip ($\mathcal{F}$), another weak measurement
($M_{3}$), and trit-flip ($\mathcal{F}$). The trit-flip operation
$\mathcal{F}$ can be realized by a $\pi$ pulse applied on the
transition $|1\rangle\leftrightarrow|2\rangle$ and followed by
another $\pi$ pulse to interchange the populations between
$|0\rangle$ and $|1\rangle$. (i.e., by the series of two $\pi$
pulses
$\pi^{|1\rangle\leftrightarrow|2\rangle}\pi^{|0\rangle\leftrightarrow|1\rangle}$)
\cite{das}.

\section{\label{s5}Conclusions}
In conclusion, we have demonstrated that weak measurement reversal
can indeed be useful for combating amplitude damping
decoherence and recovering the entanglement of two qutrits. In
particular, we have examined two simple schemes: one is
``\emph{amplitude damping} + \emph{weak measurement reversal}'' and
the other is ``\emph{weak measurement} + \emph{amplitude damping} +
\emph{weak measurement reversal}''. We have shown that the first
scheme can partially recover qutrit-qutrit entanglement for certain
initial states, but it has some limitations with respect to
entanglement protection efficiency and ESD circumvention. For the second scheme,
in which prior weak measurements and post weak measurement reversals
are carried out sequentially, the amplitude damping decoherence can
be completely suppressed for any initial states even if ESD occurs.
Even though the method is risky (i.e., a stronger procedure is
required for a longer preservation, which decreases the probability
of success), this procedure for entanglement preservation is useful
in entanglement distillation protocols and some quantum
communication tasks.

\acknowledgments

We thank the copy editor for proofreading our manuscript. This work
is supported by the Special Funds of the National Natural Science
Foundation of China under Grant Nos. 11247006 and 11247207, and by
the Natural Science Foundation of Jiangxi under Grant Nos.
20132BAB212008 and by the Scientific Research Foundation of the
Jiangxi Provincial Education Department under Grants No. GJJ13651.

\bibliography{apssamp}

\end{document}